\documentclass[11pt]{article}
\usepackage{amsmath,dsfont}
\usepackage{amsthm,amssymb}

\newcommand\version{September 24, 2007}
\numberwithin{equation}{section}

\newcommand\C{{\mathbb C}}
\newcommand\R{{\mathbb R}}
\newcommand\eps{\varepsilon}
\newcommand\half{\mbox{$\frac 12$}}

\newcommand{\B}{\mathcal{B}}

\renewcommand\epsilon\varepsilon
\renewcommand\kappa\varkappa
\renewcommand\rho\varrho
\newcommand{\x}{{\bf x}}
\newcommand{\y}{{\bf y}}
\newcommand\id{{\ensuremath {\mathds 1} }}

\newcommand\hod{{H_{\rm 1d}^{n,\ell,g}}}
\newcommand\eod{{E_{\rm 1d}^k(n,\ell,g)}}
\newcommand\htd{{H_{\rm 3d}^{n,r,\ell,a}}}
\newcommand\etd{{E_{\rm 3d}^k(n,r,\ell,a)}}

\newtheorem{thm}{THEOREM}

\newtheorem{lem}{LEMMA}
\newtheorem{cor}{COROLLARY}

\begin{document}

\markboth{\scriptsize{\version}}{\scriptsize{\version}} \title{\bf{The
    Lieb-Liniger Model as a Limit of \\ Dilute Bosons in Three
    Dimensions}} \author{\vspace{5pt} Robert Seiringer and Jun Yin \\
  \vspace{-2pt}\small{ Department of Physics, Jadwin Hall, Princeton
    University }\\ \vspace{-2pt}\small Princeton, NJ 08542-0708, USA
  \\ {\small Email: \texttt {\{rseiring/junyin\}@princeton.edu}} }

\date{\small \version}
\maketitle

\begin{abstract}
  We show that the Lieb-Liniger model for one-dimensional bosons with
  repulsive $\delta$-function interaction can be rigorously derived
  via a scaling limit from a dilute three-dimensional Bose gas with
  arbitrary repulsive interaction potential of finite scattering
  length. For this purpose, we prove bounds on both the eigenvalues
  and corresponding eigenfunctions of three-dimensional bosons in
  strongly elongated traps and relate them to the corresponding
  quantities in the Lieb-Liniger model. In particular, if both the
  scattering length $a$ and the radius $r$ of the cylindrical trap go
  to zero, the Lieb-Liniger model with coupling constant $g \sim
  a/r^2$ is derived. Our bounds are uniform in $g$ in the whole
  parameter range $0\leq g\leq \infty$, and apply to the Hamiltonian
  for three-dimensional bosons in a spectral window of size $\sim
  r^{-2}$ above the ground state energy.
\end{abstract}

\renewcommand{\thefootnote}{${\,}$}
\footnotetext{\copyright\,2007 by the authors.
This paper may be reproduced, in its entirety, for non-commercial
purposes.}

\section{Introduction}

Given the success of the Lieb-Liniger model \cite{LL,L}, both as a toy
model in statistical mechanics and as a concrete model of dilute
atomic gases in strongly elongated traps, it is worth investigating
rigorously its connection to three-dimensional models with genuine
particle interactions. A first step in this direction was taken in
\cite{bose1d,bose1da}, where it was shown that in an appropriate
scaling limit the ground state energy of a dilute three-dimensional
Bose gas is given by the ground state energy of the Lieb-Liniger
model. The purpose of this paper is to extend this result to excited
energy eigenvalues and the corresponding eigenfunctions.

The Lieb-Liniger model has recently received a lot of attention as a
model for dilute Bose gases in strongly elongated traps
\cite{olshanii,petrov1,dunjko,jackson,bergeman,bose1da,olshaniid,petrov2,bloch}.
Originally introduced as a toy model of a quantum many-body system, it
has now become relevant for the description of actual quasi
one-dimensional systems.  The recent advances in experimental
techniques have made it possible to create such quasi one-dimensional
systems in the laboratory
\cite{dettmer,moritz1,richard,kinoshita,tolra,moritz2,esteve}. These
provide a unique setting for studying matter under extreme conditions
where quantum effects dominate.

The Lieb-Liniger model describes $n$ non-relativistic bosons in
one spatial dimension, interacting via a $\delta$-function potential
with strength $
g\geq 0$. In appropriate units, the Hamiltonian is
given by 
\begin{equation}\label{defll}
  \hod=\sum_{i=1}^n\left(-\partial_i^2 + \ell^{-2}
    V^\parallel(z_i/\ell)\right)
+ g \sum_{1\leq i<j\leq n}\delta(z_i-z_j)\,.
\end{equation}
Here, we use the notation $\partial_i = \partial/\partial z_i$ for
brevity. The trap is represented by the potential $V^\parallel$, which
is assumed to be locally bounded and tend to infinity as $|z|\to
\infty$. The scaling parameter $\ell$ is a measure of the size of the
trap. Instead of the trap potential $V^\parallel$, one can also
confine the system to an interval of length $\ell$ with appropriate
boundary conditions; in fact, periodic boundary conditions were
considered in \cite{LL,L}.

The Lieb-Liniger Hamiltonian $\hod$ acts on totally symmetric wavefunctions
$\phi\in L^2(\R^{n})$, i.e., square-integrable functions satisfying
$\phi(z_1,\dots,z_n) = \phi(z_{\pi(1)},\dots,z_{\pi(n)})$ for any
permutation $\pi$. In the following, all wavefunctions will be
considered symmetric unless specified otherwise.

In the case of periodic boundary conditions on the interval
$[0,\ell]$, Lieb and Liniger have shown that the spectrum and
corresponding eigenfunctions of $\hod$ can be obtained via the Bethe
ansatz \cite{LL}. In \cite{L} Lieb has specifically studied the
excitation spectrum, which has an interesting two-branch structure.
This structure has recently received a lot of attention
\cite{jackson,bloch} in the physics literature.  Our results below
show that the excitation spectrum of the Lieb-Liniger model is a
genuine property of dilute three-dimensional bosons in strongly
elongated traps in an appropriate parameter regime. %, as long as
                                                    %$r/\ell$ and
                                                    %$a/r$ are small
                                                    %enough.

In the following, we shall consider dilute three-dimensional Bose
gases in strongly elongated traps. Here, {\it dilute} means that $a
\rho^{1/3} \ll 1$, where $a$ is the scattering length of the
interaction potential, and $\rho$ is the average particle density.
{\it Strongly elongated} means that $r \ll \ell$, where $r$ is the length
scale of confinement in the directions perpendicular to $z$. We shall
show that, for fixed $n$ and $\ell$, the spectrum of a three
dimensional Bose gas in an energy interval of size $\sim r^{-2}$ above
the ground state energy is approximately equal to the spectrum of the
Lieb-Liniger model (\ref{defll}) as long as $a \ll r$ and $r \ll
\ell$. The effective coupling parameter $g$ in (\ref{defll}) is of the
order $g\sim a/r^2$ and can take any value in $[0,\infty]$. The same
result applies to the corresponding eigenfunctions. They are
approximately given by the corresponding eigenfunctions of the
Lieb-Liniger Hamiltonian, multiplied by a product function of the
variables orthogonal to $z$. The precise statement of our results will
be given in the next section.

We remark that the problem considered here is somewhat analogous to
the one-dimensional behavior of atoms in extremely strong magnetic
fields, where the Coulomb interaction behaves like an effective
one-dimensional $\delta$-potential when the magnetic field shrinks the
cyclotron radius of the electrons to zero. For such systems, the
asymptotics of the ground state energy was studied in \cite{BSY}, and
later the excitation spectrum and corresponding eigenfunctions were
investigated in \cite{BD}. In this case, the effective one-dimensional
potential can be obtained formally by integrating out the variables
transverse to the magnetic field in a suitably scaled Coulomb
potential. Our case considered here is much more complicated, however. The
correct one-dimensional physics emerges only if the kinetic and
potential parts of the Hamiltonian are considered together.

\section{Main Results}

Consider the Hamiltonian for $n$ spinless bosons in three space
dimensions, interacting via a pair-potential $v$. We shall write
$\x = (\x^\perp,z) \in \R^3$, with $\x^\perp \in \R^2$. 
Let $V^\perp \in L_{\rm loc}^\infty(\R^2)$ and
$V^\parallel\in L_{\rm loc}^\infty(\R)$ denote the (real-valued) confining
potentials in $\x^\perp$ in $z$-direction, respectively. 
Then
\begin{equation}
  \htd=\sum_{i=1}^{n}\left(-\Delta_i+ r^{-2} V^\perp(\x_i^{\perp}/r)
    + \ell^{-2} V^\parallel(z_i/\ell)\right) 
+\sum_{1\leq i<j\leq n} a^{-2} v(|\x_i-\x_j|/a)\,.
\end{equation}
The trap potentials $V^\parallel$ and $V^\perp$ confine the motion in
the longitudinal ($z$) and the transversal ($\x^{\perp}$) directions,
respectively, and are assumed to be locally bounded and tend to
$\infty$ as $|z|$ and $|\x^{\perp}|$ tend to $\infty$. Without loss of
generality, we can assume that $V^\parallel \geq 0$. The scaling
parameters $r$ and $\ell$ measure the size of the traps.

The interaction potential $v$ is assumed to be nonnegative, of finite
range $R_0$ and have scattering length 1; the scaled potential $a^{-2}
v(|\,\cdot\,|/a)$ then has scattering length $a$ \cite{LSSY,lsy1}
and range $a R_0$. We do not assume any smoothness or even
integrability of $v$. In particular, we allow $v$ to take the value
$+\infty$ on a set of non-zero measure, corresponding to hard-sphere
particles.

Let $e^\perp$ and $b(\x^\perp)$ denote the ground state energy and the
normalized ground state wave function of
$-\Delta^\perp+V^\perp(\x^\perp)$, respectively.  Note that $b$ is a
bounded and strictly positive function and, in particular, $b\in
L^p(\R^2)$ for any $2\leq p \leq \infty$. Let also $\widetilde e^\perp>0$
denote the gap above the ground state energy of
$-\Delta^\perp+V^\perp(\x^\perp)$. The corresponding quantities for
$-\Delta^\perp+r^{-2} V^\perp(\x^\perp/r)$ are then given by $e^\perp/r^2$,
$\widetilde e^\perp/r^2$ and $b_r(\x^\perp) = r^{-1} b(\x^\perp/r)$,
respectively.

The eigenvalues of $\htd$ will be denoted by $\etd$, with
$k=1,2,3,\dots$. Moreover, the eigenvalues of $\hod$ in (\ref{defll})
will be denoted by $\eod$. Theorem~1 shows that $\etd$ is
approximately equal to $\eod + e^\perp/r^2$ for small $a/r$ and
$r/\ell$, for an appropriate value of the parameter $g$. In fact, $g$
turns out to be given by
\begin{equation}\label{defgc}
g=\frac {8\pi a}{r^2} \int_{\R^2} |b(\x^\perp)|^4 d^2\x^\perp\,.
\end{equation}

\bigskip

\begin{thm}\label{thm1}
  Let $g$ be given in (\ref{defgc}). There exist constants $C>0$
  and $D>0$, independent of $a$, $r$, $\ell$, $n$ and $k$, such that
  the following bounds hold: 
\begin{equation}\label{lowthm1}
{\rm (a)} \quad \etd \geq \frac {ne^\perp}{r^2} +  
\eod \left(1 - \eta_{\rm L}\right)\left(1- \frac{r^2}{\widetilde e^\perp}
\eod \right)
\end{equation}
as long as $\eod \leq \widetilde e^\perp/r^2$, with
\begin{equation}\label{defel}
\eta_{\rm L} = D \left[\left(
    \frac{na}{r}\right)^{1/8} + n^2 \left(
    \frac{na}{r}\right)^{3/8}\right]
\end{equation}
\begin{equation}\label{upthm1}
\!\!\!\!\!\!\!\!\!\!\!\!\!\!\!\!\!\!\!\!\!\!\!\!\!\!\!\!
\!\!\!\!\!\!\!\!\!\!\!\!\!\!\!\!\!\!\!\!\!\!\!\!\!\!\!
{\rm (b)} \qquad \etd \leq \frac {ne^\perp}{r^2} + 
\eod \left( 1 - \eta_{\rm U}\right)^{-1}
\end{equation}
whenever $\eta_{\rm U}<1$, where 
\begin{equation}\label{defeu}
\eta_{\rm U} =  C \left( \frac {na} r\right)^{2/3}\,.
\end{equation}
\end{thm}

We note that $\eod$ is monotone increasing in $g$, and uniformly
bounded in $g$ for fixed $k$. In fact, $\eod \leq E^k_{\rm f}(n,\ell)$
for all $g\geq 0$ and for all $k$, where $E^k_{\rm f}(n,\ell)$ are the
eigenenergies of $n$ non-interacting fermions in one dimension
\cite{girardeau}. Using this property of uniform boundedness, we can
obtain the following Corollary from Theorem~\ref{thm1}.

\begin{cor}\label{cor1}
Fix $k$, $n$ and $\ell$. If $r\to 0$, $a\to 0$ in such a way that
$a/r\to 0$, then
\begin{equation}
\lim \frac{ \etd - ne^\perp /r^2}{\eod} = 1 \,.
\end{equation}
\end{cor}

Note that by simple scaling $\etd = \ell^{-2} E_{\rm
  3d}^k(n,r/\ell,1,a/\ell)$, and likewise $\eod = \ell^{-2} E_{\rm
  1d}^k(n,1,\ell g)$. Hence it is no restriction to fix $\ell$ in the
limit considered in Corollary~\ref{cor1}. In fact, $\ell$ could be set
equal to $1$ without loss of generality.
Note also that the convergence stated in Corollary \ref{cor1} is uniform in $g$, in the
sense that $g\sim a r^{-2}$ is allowed to go to $+\infty$ as $r\to 0$, $a\to 0$,
as long as $g r \sim a/r \to 0$.  

In order to state our results on the corresponding eigenfunctions of
$\htd$ and $\hod$, we first have to introduce some additional notation
to take into account the possible degeneracies of the
eigenvalues. Let $k_1=1$, and let $k_i$ be recursively defined by
\[ k_i = \min \left\{ k \, :\, \eod > E_{\rm
    1d}^{k_{i-1}}(n,\ell,g)\right\}\,.
\]
Then $E_{\rm 1d}^{k_{i}}(n,\ell,g)<E_{\rm 1d}^{k_{i+1}}(n,\ell,g)$,
while $E_{\rm 1d}^{k_{i}+j}(n,\ell,g) = E_{\rm 1d}^{k_{i}}(n,\ell,g)$
for $0\leq j< k_{i+1}-k_i$.  That is, $k$ counts the energy levels
including multiplicities, while $i$ counts the levels without
multiplicities. Hence, if $k_{i} \leq k < k_{i+1}$, the energy
eigenvalue $\eod$ is $k_{i+1}-k_i$ fold degenerate.  Note that $k_i$
depends on $n$, $\ell$ and $g$, of course, but we suppress this
dependence in the notation for simplicity.

Our main result concerning the eigenfunctions of $\htd$ is as follows.

\begin{thm}\label{thm2}
Let $g$ be  given in (\ref{defgc}). Let $\Psi_k$ be an
eigenfunction of $\htd$ with eigenvalue $\etd$, and let $\psi_l$ ($k_i
\leq l < k_{i+1}$) be orthonormal eigenfunctions of $\hod$
corresponding to eigenvalue $\eod$. 
Then %, for $k_i \leq k < k_{i+1}$,
\begin{align} \nonumber &\sum_{l=k_i}^{k_{i+1}-1} \left| \left\langle
      \Psi_k \left| \psi_l \prod_{i=1}^n b_r(\x^\perp_k)\right.
    \right\rangle\right|^2 \geq 1 - \left(\frac
    {\left(1-(r^2/\widetilde
        e^\perp)\eod\right)^{-1}}{(1-\eta_{\rm U})(1-\eta_{\rm L})}-1\right) \times \\
  & \qquad \qquad \times \left[ \frac{\sum_{i=1}^{k_{i+1}-1}E_{\rm
        1d}^i(n,\ell,g)}{E_{\rm 1d}^{k_{i+1}}(n,\ell,g)-E_{\rm
        1d}^{k_i}(n,\ell,g)}+\frac{\sum_{i=1}^{k_i-1}E_{\rm
        1d}^i(n,\ell,g)}{E_{\rm 1d}^{k_{i}}(n,\ell,g)-E_{\rm
        1d}^{k_i-1}(n,\ell,g)}\right]
\end{align}
as long as $\eta_{\rm U}<1$, $\eta_{\rm L}<1$ and $\eod< \widetilde e^\perp/r^2$.
\end{thm}

In particular, this shows that $\Psi_k(\x_1,\dots,\x_n)$ is
approximately of the product form $\psi_k(z_1,\dots,z_n) \prod_i
b_r(\x_i^\perp)$ for small $r/\ell$ and $a/r$, where $\psi_k$ is an
eigenfunction of $\hod$ with eigenvalue $\eod$. Note that although
$\Psi_k$ is close to such a product in $L^2(\R^{3n})$ sense, it is
certainly not close to a product in a stronger norm involving
the energy.  In fact, a product wavefunction will have infinite energy
if the interaction potential $v$ contains a hard core; in any case, its
energy will be too big and not related to the scattering length of $v$
at all.

\begin{cor}\label{cor2}
  Fix $k$, $n$, $\ell$ and $g_0\geq 0$. Let $P_{g_0,{\rm 1d}}^k$ denote the
  projection onto the eigenspace of $H_{\rm 1d}^{n,\ell,g_0}$ with
  eigenvalue $E^k_{\rm 1d}(n,\ell,g_0)$, and
  let $P^\perp_r$ denote the projection onto the function
  $\prod_{i=1}^n b_r(\x^\perp_i) \in L^2(\R^{2n})$. If $r\to 0$, $a\to
  0$ in such a way that $g=8\pi \|b\|_4^4 a/r^2 \to g_0$, then
\begin{equation}
\lim \, \langle \Psi_k | P^\perp_r \otimes P_{g_0,{\rm 1d}}^k | \Psi_k\rangle = 1\,.
\end{equation}
\end{cor}

Here, the tensor product refers to the decomposition $L^2(\R^{3n})=
L^2(\R^{2n})\otimes L^2(\R^n)$ into the transversal ($\x^\perp$) and
longitudinal ($z$) variables.

We note that Corollary~\ref{cor2} holds also in case $g_0=\infty$. In
this case, $P_{g_0,{\rm 1d}}^k$ has to be defined as the spectral
projection with respect to the limiting energies $\lim_{g\to\infty}
\eod$.  Using compactness, it is in fact easy to see that the limit of
$P_{g_0,{\rm 1d}}^k$ as $g_0\to \infty$ exists in the operator norm
topology. We omit the details.

Our results in particular imply that $\htd$ converges to $\hod$ in a
certain norm-resolvent sense.  In fact, if $a\to 0$ and $r\to 0$ in
such a way that $g=8\pi \|b\|_4^4 a/r^2 \to g_0$, and $\lambda \in \C
\setminus [E^1_{\rm 1d}(n,\ell,g_0),\infty)$ is fixed, then it follows
easily from Corollaries~\ref{cor1} and~\ref{cor2} that
$$
\lim \, \left\| \frac 1{\lambda + ne^\perp/r^2 - \htd } - P^\perp_r
  \otimes \frac 1{\lambda - H_{\rm
      1d}^{n,\ell,g_0}} \right\| =0\,.
$$

\bigskip
%\subsection{Extensions}

Our main results, Theorems~\ref{thm1} and~\ref{thm2}, can be extended
in several ways, as we will explain now. For simplicity and
transparency, we shall not formulate the proofs in the most general
setting.

\begin{itemize}

\item Instead of allowing the whole of $\R^2$ as the configuration
  space for the $\x^\perp$ variables, one could restrict it to a
  subset, with appropriate boundary conditions for the Laplacian
  $\Delta^\perp$. For instance, if $V^\perp$ is zero and the motion is
  restricted to a disk with Dirichlet boundary conditions, the
  corresponding ground state function $b$ in the transversal
  directions is given by a Bessel function.

\item Similarly, instead of taking $\R$ as the configuration space for
  the $z$ variables, one can work on an interval with appropriate
  boundary conditions. In particular, the case $V^\parallel = 0$ with
  periodic boundary conditions on $[0,\ell]$ can be considered, which
  is the special case studied by Lieb and Liniger in \cite{LL,L}.

\item As noted in previous works on dilute Bose gases \cite{lsy1,LSSY}, the
  restriction of $v$ having a finite range can be
  dropped. Corollaries~\ref{cor1} and~\ref{cor2} remain true for all
  repulsive interaction potentials $v$ with finite scattering
  length. Also Theorems~\ref{thm1} and~\ref{thm2} remain valid, with
  possibly modified error terms, however, depending on the rate of
  decay of $v$ at infinity. We refer to \cite{lsy1,LSSY} for details.

\item Our results can be extended to any symmetry type of the 1D
  wavefunctions, not just symmetric ones. In particular, one can allow
  the particles to have internal degrees of freedom, like spin. 

\item As mentioned in the Introduction, in the special case $k=1$,
  i.e., for the ground state energy, similar bounds as in Theorem 1
  have been obtained in \cite{bose1d}. In spite of the fact that the
  error terms are not uniform in the particle number $n$, these bounds
  have then been used to estimate the ground state energy in the
  thermodynamic limit, using the technique of Dirichlet--Neumann
  bracketing. Combining this technique with the results of Theorem 1,
  one can obtain appropriate bounds on the free energy at positive
  temperature and other thermodynamic potentials in the thermodynamic
  limit as well. In fact, since our energy bounds apply to all
  (low-lying) energy eigenvalues, bounds on the free energy in a
  finite volume are readily obtained. The technique employed in
  \cite{bose1d} then allows for an extension of these bounds to
  infinite volume (at fixed particle density).

\end{itemize}

%\subsection{Outline}

In the following, we shall give the proof of Theorems~\ref{thm1}
and~\ref{thm2}. The next Section~\ref{sect:up} gives the proof of the
upper bound to the energies $\etd$, as stated in Theorem~\ref{thm1}(b).
The corresponding lower bounds in Theorem~\ref{thm1}(a) are proved in
Section~\ref{sect:low}. Finally,  the proof of
Theorem~\ref{thm2} will be given
in Section~\ref{sect:ef}. 

\section{Upper Bounds}\label{sect:up}

This section contains the proof of the upper bounds to the 3D energies
$\etd$, stated in Theorem~\ref{thm1}(b). Our strategy is similar to
the one in \cite[Sect.~3.1]{bose1d} and we will use some of the
estimates derived there. The main improvements presented here concern
the extension to excited energy eigenvalues, and the derivation of a
bound that is uniform in the effective coupling constant $g$ in
(\ref{defgc}). In contrast, the upper bound in \cite[Thm.~3.1]{bose1d}
applies only to the ground state energy $E_{\rm 3d}^1(n,r,\ell,a)$,
and is not uniform in $g$ for large $g$.

Before proving the upper bounds to the 3D energies $\etd$, we will
prove a simple lemma that will turn out to be useful in the following.

\begin{lem}\label{lem1}
  Let $H$ be a non-negative Hamiltonian on a Hilbertspace
  $\mathcal H$, with eigenvalues $0\leq E_1\leq E_2 \leq E_3 \leq
  \dots$. For $k\geq 1$, let
  $f_1,\dots ,f_k \in \mathcal H$. If, for any $\{a_i\}$ with
  $\sum_{i=1}^k |a_i|^2=1$ we have
  \[ \left\|\mbox{$\sum_{i=1}^k$} a_if_i\right\|^2\geq 1-\epsilon
  \quad {\it for\ } \epsilon<1\] and
\[ \left\langle \mbox{$\sum_{i=1}^k$} a_if_i\right|H\left|
  \mbox{$\sum_{i=1}^k$} a_if_i\right\rangle\leq E \,,\] then $E_k\leq
E(1-\epsilon)^{-1}$.
\end{lem}

\begin{proof}
  For any $f$ in the $k$-dimensional subspace spanned by the $f_i$, we
  have
\[\langle f|H|f\rangle\leq E(1-\epsilon)^{-1}\langle f|f\rangle \]
by assumption. Hence $E_k\leq E/(1-\epsilon)$ by the variational principle.
\end{proof}

Pick $\{a_i\}$ with $\sum_{i=1}^k |a_i|^2 = 1$, and let $\phi =
\sum_{i=1}^k a_i \psi_i$, where the $\psi_i$ are orthonormal
eigenfunctions of $\hod$ with eigenvalues $E_{\rm 1d}^i(n,\ell,g)$.
Consider a 3D trial wave function of the form
\[\Phi(\x_1,\dots,\x_n
) = \phi(z_1,\dots,z_n) F(\x_1,\dots,\x_n) \prod_{k=1}^n
b_r(\x_k^\perp) \,, \]
where $F$ is defined by
\[F(\x_1,\ldots,\x_n)=\prod_{i<j}f(|\x_i-\x_j|) \,.\] Here, $f$ is a
function with $0\leq f\leq 1$, monotone increasing, such that $f(t)=1$
for $t\geq R$ for some $R\geq a R_0$. For $t\leq R$ we shall choose
$f(t)=f_0(t)/f_0(R)$, where $f_0$ is the solution to the zero-energy
scattering equation for $a^{-2} v(|\,\cdot\,|/a)$
\cite{LSSY,lsy1}. That is, 
$$
\left( -\frac {d^2}{dt^2} - \frac 2 t \frac{d}{dt} + \frac 1{2a^2}
  v(t/a) \right) f_0(t) = 0\,,
$$ 
normalized such that $\lim_{t\to\infty} f_0(t)=1$. 
This $f_0$ has the properties that $f_0(t)=1-a/t$ for $t\geq a R_0$, and
$f'_0(t)\leq t^{-1}\min\{1,a/t\}$.

To be able to apply Lemma~\ref{lem1}, we need a lower bound on the
norm of $\Phi$. This can be obtained in the same way as in
\cite[Eq.~(3.9)]{bose1d}.  Let $\rho^{(2)}_{\phi}$ denote the two-particle
density of $\phi$, normalized as
 \[\int_{\R^2} \rho^{(2)}_{\phi}(z,z')dzdz'=1 \,.\] 
Since $F$ is 1 if no pair of particles is closer together than a distance
$R$, we can estimate the norm of $\Phi$ by 
\begin{align}\nonumber
\langle\Phi|\Phi\rangle & \geq  
1-\frac{n(n-1)}{2} \int_{\R^6}
\rho^{(2)}_{\phi}(z,z')b_r(\x^\perp)^2b_r(\y^\perp)^2\theta(R-|\x-\y|)
dzdz'd^2\x^\perp d^2\y^\perp \\ \nonumber &\geq  1-\frac{n(n-1)}{2}
\int_{\R^4} b_r(\x^\perp)^2b_r(\y^\perp)^2\theta(R-|\x^\perp-\y^\perp|)
d^2\x^\perp d^2\y^\perp \\  &\geq 1- \frac{n(n-1)}2
\frac {\pi R^2}{r^2} \|b\|_4^4 \label{correct}\,,
\end{align}
where Young's inequality \cite[Thm.~4.2]{anal} has been used in the
last step. Since $F\leq 1$, the norm of $\Phi$ is less than 1, and hence
\begin{equation}\label{norm1}
1\geq \langle\Phi|\Phi\rangle \geq 1-\frac{n(n-1)}2
\frac {\pi R^2}{r^2} \|b\|_4^4 \,. 
\end{equation}

Next, we will derive an upper bound on $\langle \Phi|
\htd|\Phi\rangle$. 
Define $G$ by $\Phi = G F$. Using partial integration and the fact
that $F$ is real-valued, we have 
\[
\langle\Phi|-\Delta_j|\Phi\rangle=-\int_{\R^{3n}} F^2 \overline G\Delta_j G +
\int_{\R^{3n}} |G|^2 |\nabla_j F|^2 \,.  
\]
Using $-\Delta^\perp b_r =
(e^\perp/r^2)b_r - r^{-2} V^\perp(\,\cdot\,/r) b_r$, we therefore get
\begin{align}\nonumber 
\langle\Phi|\htd|\Phi\rangle &=
\frac{ne^\perp}{r^2}\langle\Phi|\Phi\rangle + \int_{\R^{3n}} F^2 \prod_{j=1}^n
b_r(\x^\perp_j)^2 \, \overline\phi \,\big( H_{\rm 1d}^{n,\ell,g}
\phi\big) \\\nonumber &\quad - g \left\langle\Phi\left|
    \mbox{$\sum_{i<j}$}\delta(z_i-z_j)\right|\Phi\right\rangle \\
\label{exp} &\quad  + \int_{\R^{3n}} |G|^2 \left( \sum_{j=1}^n |\nabla_j F|^2 +
  \sum_{i<j} a^{-2} v(|\x_i-\x_j|/a)|F|^2\right) \, .
\end{align}

With the aid of Schwarz's inequality for the integration over the
$z$ variables, as well as $F\leq 1$,
\begin{align}\nonumber
  &\int_{\R^{3n}} F^2 \prod_{j=1}^n b_r(\x^\perp_j)^2 \, \overline \phi
  \big( \hod\phi\big)\\\nonumber
  &\leq \left(\int_{\R^{3n}} F^2 \prod_{j=1}^n b_r(\x^\perp_j)^2 |\phi|^
2\right)^{1/2}\left(\int_{\R^{3n}} F^2 \prod_{j=1}^n b_r(\x^\perp)^2 
\big|\hod \phi\big|^2\right)^{1/2} \\
  &\leq \|\phi\|_2\|\hod\phi\|_2 \leq \eod\,. \label{2p5}
\end{align}
Here, we have used that $\phi$ is a linear combination of the
first $k$ eigenfunctions of $\hod$ to obtain the final inequality.
The term in the second line of (\ref{exp}) is bounded by
\begin{equation}\label{cha3} 
\left\langle\Phi\left|
    \mbox{$\sum_{i<j}$}\delta(z_i-z_j)\right|\Phi\right\rangle \geq
\frac{n(n-1)}{2} \int_\R \rho_\phi^{(2)}(z,z) dz
\left(1-\frac{n(n-1)}{2}\frac{\pi R^2}{r^2} \|b\|_4^4\right) \,,
\end{equation}
as an argument similar to (\ref{correct}) shows. 

The remaining last term in (\ref{exp}) can be bounded in the similar
way as in \cite[Sect.~3]{bose1d}. In fact, by repeating the analysis
in \cite[Eqs.~(3.12)--(3.19)]{bose1d}
\begin{align}\nonumber
  &\int_{\R^{3n}} |G|^2 \left( \sum_{j=1}^n |\nabla_j F|^2 + \sum_{i<j} a^{-2}
    v(|\x_i-\x_j|/a)|F|^2\right) \\ \nonumber &\leq \frac{n(n-1)}{r^2}
  \|b\|_4^4 \int_{\R^2} \rho^{(2)}_\phi(z,z') h(z-z') dz dz' \\ &
  \quad + \frac {2}{3}
  n(n-1)(n-2)\frac{\|b\|_\infty^2}{r^2}\frac{\|b\|_4^4}{r^2}\|m\|_\infty\int_{\R^2}
  \rho^{(2)}_\phi(z,z') m(z-z') dz dz' \,. \label{2p7}
\end{align}
Here, 
\[
h(z)=\int_{\R^2} \left( f'(|\x|)^2+\half a^{-2} v(|\x|/a)
f(|\x|)^2\right)d^2\x^\perp 
\]
and
\[
m(z)=\int_{\R^2} f'(|\x|)d^2\x^\perp \, .
\]
(The function $m$ was called $k$ in \cite{bose1d}.) 
Note that both $h$ and $m$ are supported in $[-R,R]$, and $\int_\R h(z)dz= 4\pi a (1 -
a/R)^{-1}$. 

We now proceed differently than in \cite{bose1d}. Our analysis here
has the advantage of yielding an upper bound that is uniform in $g$.
Let $\varphi\in H^1(\R)$. Then
\begin{align}\nonumber 
\left| |\varphi(z)|^2-|\varphi(z')|^2\right| &= \left| \int_{z'}^z \frac
{d|\varphi(t)|^2}{dt} dt\right| \\  \label{39}  & \leq 2 |\varphi(z')| \int_{z'}^z
\left|\frac {d\varphi(t)}{dt}\right| dt+2\left(\int_{z'}^z
\left|\frac {d\varphi(t)}{dt}\right| dt\right)^2
\\ \nonumber &\leq 2 |\varphi(z')||z-z'|^{1/2} \left(\int_{\R} \left|\frac
{d\varphi}{dz}\right|^2\right)^{1/2}+ 2|z-z'| \left(\int_{\R} \left|\frac
{d\varphi}{dz}\right|^2\right)\,.
\end{align}
Here, we used $|\varphi(t)|\leq |\varphi(z')|+\int_{z'}^z \left|\frac
  {d\varphi(t)}{dt}\right| dt$ for $z'\leq t\leq z$ for the first
inequality and applied Schwarz's inequality for the second.

We apply the bound (\ref{39}) to $\rho_\phi^{(2)}(z,z')$ for fixed
$z'$. Using the fact that the support of $h$ is contained in $[-R,R]$,
we get 
\begin{align}\nonumber 
  &\int_{\R^2} \rho^{(2)}_\phi(z,z') h(z-z') dz dz'- \int_\R h(z) dz
  \int_\R \rho_\phi^{(2)}(z,z)dz \\ \nonumber & \leq 2 R^{1/2} \int_\R
  h(z) dz \int_\R \left(\int_\R\left |\partial_z
      \sqrt{\rho_\phi^{(2)}(z,z')}\right|^2
    dz\right)^{1/2}\sqrt{\rho_\phi^{(2)}(z',z')} \, dz' \\ \nonumber &
  \quad + 2R \int_\R h(z) dz \int_{\R^2} \left|\partial_z
    \sqrt{\rho_\phi^{(2)}(z,z')}\right|^2 dz dz' \\
  &\leq \frac 2 n\, \eod \int_\R h(z)
  dz\left[\left(\frac{2R}{(n-1)g}\right)^{1/2}+ R\right]\,. \nonumber
\end{align}
Here, we used Schwarz's 
inequality and the fact that $g\half n(n-1)\int_\R
\rho_\phi^{(2)}(z,z)dz\leq \eod$ and 
\[  \int_{\R^2}  \left|\partial_z
    \sqrt{\rho_\phi^{(2)}(z,z')}\right|^2 dz dz' \leq
  \left\langle\phi\left| -\frac{d^2}{dz_1^2}\right|\phi\right\rangle\leq
\eod/n \,.\]

The same argument is used with $h$ replaced by $m$. Now $\int_\R
h(z)dz= 4\pi a (1 - a/R)^{-1}$, and \cite[Eq.~(3.22)]{bose1d}
\begin{align}\nonumber
 \|m\|_\infty &\leq \frac {2\pi a}{1-a/R}
\left(1+\ln(R/a)\right) \, , 
\\ \nonumber 
\int_\R m(z) dz &\leq \frac{ 2\pi
  aR}{1-a/R}\left(1-\frac a{2R}\right) \, .
\end{align} 
Therefore
\begin{equation}\label{bb1}
 (\ref{2p7}) \leq \frac {n(n-1)}{2} g \frac
{1+K}{1-a/R}\left( \int_\R \rho_\phi^{(2)}(z,z)dz
  +\frac 2 n \left[\left(\frac{2R}{(n-1)g}\right)^{1/2}+ R\right]\eod\right)
\end{equation}
where we denoted 
\[
K=\frac{2\pi}3 (n-2) \frac aR \frac
{1+\ln(R/a)}{1-a/R}\left(\frac Rr\right)^2 \|b\|_\infty^2 \, .  
\]

Putting together the bounds (\ref{2p5}), (\ref{cha3}) and (\ref{bb1}),
and using again the fact that $g\half n(n-1)\int_\R \rho_\phi^{(2)}(z,z)dz\leq
\eod$, we obtain the upper bound 
\begin{align}\nonumber
  \left\langle\Phi\left|\htd -\frac
      {ne^\perp}{r^2}\right|\Phi\right\rangle &\leq \eod
  \Biggl(1+\frac{n(n-1)}{2}\frac{\pi R^2}{r^2}\|b\|_4^4 + \frac
  {a/R+K}{(1-a/R)} \\ &\qquad\qquad\qquad +\frac {1+K}{1-a/R}\left[(2R
    g (n-1))^{1/2}+ (n-1)R g\right]\Biggl)\,.  \nonumber
\end{align}

It remains to choose $R$. If we choose
\[
R^3 = \frac {a r^2} {n^2}
\]
then 
\[
\left\langle\Phi\left|\htd -\frac
    {ne^\perp}{r^2}\right|\Phi\right\rangle
\leq \eod  \left(1+ C \left(\frac{n
a}{r}\right)^{2/3}\right) 
\]
for some constant $C>0$.  Moreover, from (\ref{norm1}) we see that
\[
\langle\Phi|\Phi\rangle \geq 1-C \left(\frac{n
a}{r}\right)^{2/3}\,. 
\]
Hence the upper bound (\ref{upthm1}) of Theorem~\ref{thm1} follows
with the aid of Lemma~\ref{lem1}.

\section{Lower Bounds}\label{sect:low}

In this section, we will derive a lower bound on the operator $\htd$
in terms of the Lieb-Liniger Hamiltonian $\hod$. In particular, this
will prove the desired lower bounds on the energies $\etd$. Our method is based on
\cite[Thm.~3.1]{bose1d}, but extends it in several important ways which we
shall explain.

Let $\Phi$ be a normalized wavefunction in $L^2(\R^{3n})$. We define
$f\in L^2(\R^n)$ by 
\begin{equation}\label{deff}
f(z_1,\dots,z_n)=\int_{\R^{2n}} \Phi(\x_1,\dots,\x_n)\prod_{k=1}^n
b_r(\x^\perp_k) d^2\x^\perp_k\,.
\end{equation}
Moreover, we define $F$ by
\[
\Phi(\x_1,\dots,\x_n) = F(\x_1,\dots, \x_n) \prod_{k=1}^n
b_r(\x^\perp_k)\,.
\]
Note that $F$ is well-defined, since $b_r$ is a strictly positive
function. Finally, let $G$ be given by 
\begin{equation}\label{defg}
G(\x_1,\dots,\x_n) = \Phi(\x_1,\dots,\x_n) - f(z_1,\dots,z_n)\prod_{k=1}^n
b_r(\x^\perp_k)\,.
\end{equation}

Using partial integration and the eigenvalue equation for $b_r$, we obtain 
\begin{align}\label{329}
\left\langle \Phi \left| \htd -\frac{n e^\perp}{r^2}\right|
  \Phi\right \rangle  & =
\sum_{i=1}^n \int_{\R^{3n}}  \Biggl[ |\nabla_i F|^2 + \ell^{-2} V^\parallel
  (z_i/\ell)|F|^2 \\ \nonumber & \qquad\qquad  +\half  \sum_{j,\, j\neq i}
a^{-2} v(|\x_i-\x_j|/a)|F|^2 \Biggl] \prod_{k=1}^n b_r(\x^\perp_k)^2 d^3
\x_k .
\end{align}
Now choose some $R>a R_0$, and let 
\[
U(r)=\left\{\begin{array}{ll } 3(R^3-a R_0^3)^{-1} & {\rm for\
}a R_0\leq r\leq R
\\ 0 & {\rm otherwise} \ .
\end{array} \right.
\]
For $\delta>0$ define $\B_\delta\subset\R^2$ by
\[
\B_\delta=\left\{ \x^\perp\in \R^2 \, :\, b(\x^\perp)^2\geq
  \delta\right\} \ .
\]
Proceeding along the same lines as in \cite[Eqs.~(3.31)--(3.36)]{bose1d}, we
conclude that, for any $0\leq \eps\leq 1$,
\begin{align}\nonumber
  &\sum_{i=1}^n \int_{\R^{3n}} \biggl[ |\nabla_i F|^2 +\half \sum_{j,\, j\neq i}
    a^{-2} v(|\x_i-\x_j|/a)|F|^2 \biggl] \prod_{k=1}^n b_r(\x^\perp_k)^2 d^3
  \x_k \\ \nonumber & \geq \sum_{i=1}^n \int_{\R^{3n}} \biggl[ \eps
    |\nabla_i^\perp F|^2 + a'
    U(|\x_i-\x_{k(i)}|)\chi_{\B_\delta}(\x^\perp_{k(i)}/r)|F|^2\biggl]
  \prod_{k=1}^n b_r(\x^\perp_k)^2 d^3 \x_k\\ & \quad + \int_{\R^{3n}}
  \biggl[ \eps |\partial_i \Phi|^2 +(1-\eps)|\partial_i \Phi|^2
    \chi_{\min_k |\x_i-\x_k|\geq R}(\x_i) \biggl]\prod_{k=1}^n d^3 \x_k \,.
    \label{negl}
\end{align}
Here, $\x_{k(i)}$ denote the nearest neighbor of $\x_i$ among the
$\x_j$ with $j\neq i$, $\x_{k(i)}^\perp$ is its $\perp$-component, and
$a'=a(1-\eps)(1-2 R \|\nabla b^2\|_\infty/ r \delta)$. (It is easy to
see that $\nabla b^2$ is a bounded function; see, e.g., the proof of
Lemma~1 in the Appendix in \cite{bose1d}.)  The characteristic
function of $\B_\delta$ is denoted by $\chi_{\B_\delta}$, and
$\chi_{\min_k |\x_i-\x_k|\geq R}$ restricts the $\x_i$ integration to
the complement of the balls of radius $R$ centered at the $\x_k$ for
$k\neq i$. That is, for the lower bound in (\ref{negl}) only the
kinetic energy inside these balls gets used.  (Compared with
\cite[Eq.~(3.36)]{bose1d}, we have dropped part of the kinetic energy
in the $\x^\perp$ direction in the last line, which is legitimate for
a lower bound.)

For a lower bound, the characteristic function $\chi_{\min_k
  |\x_i-\x_k|\geq R}$ could be replaced by the smaller quantity
$\chi_{\min_k |z_i-z_k|\geq R}$, as was done in \cite{bose1d}. We do 
not do this here, however, and this point will be important in the
following. In particular, it allows us to have the full kinetic
energy (in the $z$ direction) at our disposal in the effective
one-dimensional problem that is obtained after integrating out the
$\x^\perp$ variables. In contrast, only the kinetic energy in the
regions $|z_i-z_k|\geq R$ was used in \cite{bose1d} to derive a lower
bound on the ground state energy. The improved method presented in the
following leads to an operator lower bound, however.

We now give a lower bound on the two terms on the right side of
(\ref{negl}). We start with the second term, which is bounded from
below by 
\begin{align}\nonumber
 &\int_{\R^{3n}} |\partial_i \Phi|^2 \prod_{k=1}^n d^3 \x_k\\ & -
 (1-\epsilon)\int_{\R^{3n}} |\partial_i \Phi|^2 \chi_{\min_k
   |z_i-z_k|\leq R}(z_i)\chi_{\min_k |\x^\perp_i-\x^\perp_k|\leq
   R}(\x^\perp_i)\prod_{k=1}^n \nonumber 
  d^3 \x_k\,.
\end{align}
To estimate the last term from below, consider first the integral over
the $\x^\perp$ variables for fixed values of $z_1,\ldots, z_n$.
Using (\ref{defg}), this integral equals 
\begin{align}\nonumber
&\int_{\R^{2n}} \left|\partial_i f \mbox{$\prod_{k=1}^n$}
  b_r(\x^\perp_k)+\partial_i G\right |^2\chi_{\min_k
  |\x^\perp_i-\x^\perp_k|\leq R}(\x^\perp_i) \prod_{k=1}^n
d^2\x^\perp_k\\ \nonumber 
&\leq \left( 1+ \eta^{-1}\right)  |\partial_i f|^2 \int_{\R^{2n}} \chi_{\min_k
  |\x^\perp_i-\x^\perp_k|\leq R}(\x^\perp_i) \prod_{k=1}^n
b_r(\x^\perp_k)^2 d^2\x^\perp_k \\ \nonumber
& \quad + \left (1+\eta\right) \int_{\R^{2n}} |\partial_i G|^2
\prod_{k=1}^n d^2\x^\perp_k 
\end{align}
for any $\eta>0$, by Schwarz's inequality. It is easy to see that 
\[
\int_{\R^{2n}} \chi_{\min_k
  |\x^\perp_i-\x^\perp_k|\leq R}(\x^\perp_i) \prod_{k=1}^n
b_r(\x^\perp_k)^2 d^2\x^\perp_k \leq 
\frac{n(n-1)}{2}\frac{\pi R^2}{r^2}\|b\|^4_4
\]
using Young's inequality, as in (\ref{correct}).

Now 
\[
\int_{\R^{3n}} |\partial_i \Phi|^2 \prod_{k=1}^n d^3 \x_k
=  \int_{\R^{3n}} |\partial_i f|^2 \prod_{k=1}^n b_r(\x^\perp_k)^2 d^3 \x_k
+ \int_{\R^{3n}} |\partial_i G|^2 \prod_{k=1}^n d^3 \x_k\,,
\]
since, by the definition of $G$ in (\ref{defg}), $\partial_i G$ is
orthogonal to any function of the form $\xi(z_1,\dots,z_n)\prod_{k}
b_r(\x_k^\perp)$. We have thus shown the the second term on the right
side of (\ref{negl}) satisfies the lower bound
\begin{align}\nonumber
  & \int_{\R^{3n}} \biggl[ \eps |\partial_i \Phi|^2
    +(1-\eps)|\partial_i \Phi|^2 \chi_{\min_k |\x_i-\x_k|\geq R}(\x_i)
  \biggl]\prod_{k=1}^n d^3 \x_k \\ \nonumber &\geq
  \left(1-(1-\epsilon)\left(1+\eta^{-1}\right)\frac{n(n-1)}{2}\frac{\pi
      R^2}{r^2}\|b\|^4_4\right) \int_{\R^n} |\partial_i f|^2
  \prod_{k=1}^n dz_k \\ & \quad + \left( 1-
    (1-\epsilon)(1+\eta)\right) \int_{\R^{3n}} |\partial_i G|^2
  \prod_{k=1}^n d^3\x_k \label{col1}
\end{align}
for any $\eta >0$. We are going to choose $\eta$ small enough such
that $(1-\epsilon)(1+\eta)\leq 1$, in which case the last term is
non-negative and can be dropped for a lower bound. Note that the right
side of (\ref{col1}) contains the full kinetic energy of the function
$f$. Had we replaced $\chi_{\min_k |\x_i-\x_k|\geq R}$ by the smaller
quantity $\chi_{\min_k |z_i-z_k|\geq R}$ on the left side of
(\ref{col1}), as in \cite{bose1d}, only the kinetic energy for
$|z_i-z_k|\geq R$ would be at our disposal. This is sufficient for a
lower bound on the ground state energy, as in \cite{bose1d}, but would
not lead to the desired operator lower bound which is derived in this
section.

We now investigate the first term on the right side of (\ref{negl}).
Consider, for fixed $z_1,\dots,z_n$, the expression
\begin{equation}\label{44}
\sum_{i=1}^n \int_{\R^{2n}} \left[ \eps|\nabla^\perp_i F|^2 + a'
U(|\x_i-\x_{k(i)}|)\chi_{\B_\delta}(\x^\perp_{k(i)}/r)|F|^2
\right] \prod_{k=1}^n b_r(\x^\perp_k)^2 d^2 \x_k^\perp \ .
\end{equation}
Proceeding as in \cite[Eqs.~(3.39)--(3.46)]{bose1d}, we conclude that
\[
(\ref{44}) \geq a'' \sum_{i<j} d(z_i-z_j) \int_{\R^{2n}}
|\Phi(\x_i,\dots,\x_n)|^2 \prod_{k=1}^n d^2\x^\perp_k\,,
\]
where
\begin{align}\nonumber
a'' & = a'  \left(1-\frac {3n}{\eps\widetilde e^\perp}
\frac{ar^2}{R^3} \frac 1{1-(a R_0/R)^3}
\left[1-\frac{n^2}{\eps\widetilde e^\perp}
\frac{a}{R}3\pi\|b\|_4^4 \frac 1{1-(a R_0/R)^3}\right]^{-1}\right) \\ &
\qquad \times \left(1-(n-2) \frac{\pi R^2}{r^2} \|b\|_\infty^2\right)
 \label{adp}
\end{align}
(which is called $a'''$ in \cite{bose1d}), and 
\[
d(z-z')=\int_{\R^4} b_r(\x^\perp)^2 b_r(\y^\perp)^2 U(|\x-\y|)
\chi_{\B_\delta}(\y^\perp/r) d^2\x^\perp d^2\y^\perp \ .
\]
Note that $d(z)=0$ if $|z|\geq R$. Recall that the quantity $\widetilde e^\perp$
in (\ref{adp}) is
the gap above the ground state energy of $-\Delta^\perp + V^\perp$,
which is strictly positive.

By using
(\ref{deff})--(\ref{defg}), 
\[
\int_{\R^{2n}} |\Phi(\x_i,\dots,\x_n)|^2 \prod_{k=1}^n
d^2\x^\perp_k \geq |f(z_1,\dots,z_n)|^2\,.
\]
Let $g'= a'' \int_\R d(z) dz$. It was shown in
\cite[Eqs.~(3.49)--(3.53)]{bose1d} that, for any $\kappa >0$ and
$\varphi\in H^1(\R)$,
\begin{align}\nonumber
& a'' \int_\R d(z-z') |\varphi(z)|^2 dz \\ & \geq g'\,
  \max_{z,\, |z-z'|\leq R} |\varphi(z)|^2 \left( 1 -
        \sqrt{\frac{2g' R}{\kappa}} \right)  - \kappa \int_\R
      |\partial \varphi|^2 dz \nonumber
\end{align}
for fixed $z'$. In particular, applying this estimate to $f$ for fixed
$z_j$, $j\neq i$, 
\begin{align}\nonumber
& a'' \sum_{i<j} \int_{\R^n} d(z_i-z_j) |f|^2 \prod_{k=1}^n dz_k \\ &\geq 
g'' \sum_{i<j}  \int_{\R^n} \delta(z_i-z_j) |f|^2 \prod_{k=1}^n dz_k
- (n-1) \kappa \sum_{i=1}^n \int_{\R^n} |\partial_i f|^2 \prod_{k=1}^n
dz_k  \,, \nonumber
\end{align}
with 
\begin{equation}\label{gpp} 
g''=g'\left(1-\sqrt{\frac {2 g'R}{\kappa}}\right)
\,.  
\end{equation} 
Hence we have shown that 
\begin{align}\nonumber
& \sum_{i=1}^n \int \biggl[ \eps
    |\nabla_i^\perp F|^2 + a'
    U(|\x_i-\x_{k(i)}|)\chi_{\B_\delta}(\x^\perp_{k(i)}/r)|F|^2\biggl]
  \prod_{k=1}^n b_r(\x^\perp_k)^2 d^3 \x_k \\ & \geq g''
\sum_{i<j}  \int_{\R^n} \delta(z_i-z_j) |f|^2 \prod_{k=1}^n dz_k
- (n-1) \kappa \sum_{i=1}^n \int_{\R^n} |\partial_i f|^2 \prod_{k=1}^n
dz_k \,.
\label{col2}
\end{align}
This concludes our bound on the first term on the right side of
(\ref{negl}).

For the term involving $V^\parallel$ in (\ref{329}), we can again use
the orthogonality properties of $G$ in (\ref{defg}), as well as the
assumed positivity of $V^\parallel$, to conclude that
\begin{align}\nonumber
  &\sum_{i=1}^n \int_{\R^{3n}} \ell^{-2} V^\parallel (z_i/\ell)
  |\Phi(\x_1,\dots,\x_n)|^2 \prod_{k=1}^n d^3 \x_k \\ &\geq
  \sum_{i=1}^n \int_{\R^{n}} \ell^{-2} V^\parallel (z_i/\ell)
  |f(z_1,\dots,z_n)|^2 \prod_{k=1}^n dz_k\,. \label{col3}
\end{align}
By combining (\ref{329}) and (\ref{negl}) with the estimates (\ref{col1}),
(\ref{col2}) and (\ref{col3}), we thus obtain
\begin{align}\nonumber
  &\left\langle \Phi \left| \htd -\frac{n e^\perp}{r^2}\right|
    \Phi\right \rangle \\ \nonumber & \geq
  \left(1-(1-\epsilon)\left(1+\eta^{-1}\right)\frac{n(n-1)}{2}\frac{\pi
      R^2}{r^2}\|b\|^4_4 - (n-1) \kappa \right) \int_{\R^n}
  |\partial_i f|^2 \prod_{k=1}^n dz_k \\ & \quad + \int_{\R^{n}}
  \left[ \sum_{i=1}^n  \ell^{-2} V^\parallel (z_i/\ell) +
    g'' \sum_{i<j}\delta(z_i-z_j)\right] |f|^2
  \prod_{k=1}^n dz_k\,. \label{fin2}
\end{align}
Recall that $g''$ is given in (\ref{gpp}). This estimate is valid for
all $0\leq \epsilon\leq 1$, $\kappa>0$ and $\eta>0$ such that
$(1-\epsilon)(1+\eta)\leq 1$ in order to be able to drop the last
term in (\ref{col1}).

To complete the estimate, it remains to give a lower bound to $\int_\R d(z) dz$. 
As in \cite[Eq.~(3.48)]{bose1d}, we can use
$|b(\x^\perp)^2-b(\y^\perp)^2|\leq R \|\nabla b^2\|_\infty$ for
$|\x^\perp-\y^\perp|\leq R$ to estimate
\begin{align}\nonumber 
\int_\R d(z) dz &\geq
\frac {4\pi}{r^2} \left(\int_{\B_\delta}
  b(\x^\perp)^4 d^2\x^\perp - R\|\nabla b^2\|_\infty/r\right)\\ \nonumber
&\geq \frac {4\pi}{r^2}\left(\|b\|_4^4- \delta- R\|\nabla
  b^2\|_\infty/r\right) \,. 
\end{align}
This leads to the bound 
\begin{align}\nonumber
  \frac{g''}{g}&\geq \left(1-\sqrt{\frac{2
        gR}{\kappa}}\right)\left(1-\frac{ \delta}{\|b\|_4^4}-
    \frac{R\|\nabla b^2\|_\infty/r}{\|b\|_4^4}\right)\\ \nonumber
  &\quad \times \left(1-\eps\right)\left(1-2 R \|\nabla b^2\|_\infty/
    r \delta\right)\left(1-(n-2) \frac{\pi R^2}{r^2}
    \|b\|_\infty^2\right)\\\nonumber &\quad \times \left(1-\frac
    {3n}{\eps\widetilde e^\perp} \frac{ar^2}{R^3} \frac 1{1-(a R_0/R)^3}
    \left[1-\frac{n^2}{\eps\widetilde e^\perp}
      \frac{a}{R}3\pi\|b\|_4^4 \frac 1{1-(a R_0/R)^3}\right]\right)\,.
\end{align}

It remains to choose the free parameters $R$, $\delta$, $\kappa$,
$\epsilon$ and $\eta$. If we choose 
\[ R = n \left( \frac{na}{r}\right)^{1/4} \quad , \quad \delta =
\epsilon = \eta = 
\left( \frac{na}{r}\right)^{1/8} \quad , \quad \kappa = \frac 1n
\left( \frac{na}{r}\right)^{5/12} \,, \]
Eq.~(\ref{fin2}) implies
\[
\left\langle \Phi \left| \htd -\frac{n e^\perp}{r^2}\right|
  \Phi\right \rangle \geq \left( 1 - D \left(
    \frac{na}{r}\right)^{1/8} - D n^2 \left(
    \frac{na}{r}\right)^{3/8}\right) 
\left\langle f \left| \hod \right| f\right\rangle
\]
for some constant $D>0$. In particular, if $P^\perp_r$ denotes the
projection onto the function $\prod_{k=1}^n b_r(\x^\perp_k) \in
L^2(\R^{2n})$, we have the operator inequality
\[
\htd - \frac{ n e^\perp}{r^2} \geq \left(1-\eta_{\rm L}\right)  P^\perp_r \otimes \hod 
\]
in the sense of quadratic forms. (Here, $\eta_{\rm L}$ is defined in
(\ref{defel}).) 

Recall that that the gap above the ground state
energy of $-\Delta^\perp + r^{-2} V^\perp(\x^\perp/r)$ is
$\widetilde e^\perp/r^2$. Hence, using the positivity of both
$V^\parallel$ and $v$, 
\[
\htd - \frac{ n e^\perp}{r^2} \geq  \frac {\widetilde e^\perp}{r^2} 
 \left(\id-P^\perp_r\right) \otimes \id\,.
\]
In particular, 
\begin{equation}\label{finb}
\htd  - \frac{ n e^\perp}{r^2} \geq (1-\gamma) \left(1-\eta_{\rm
    L}\right)  P^\perp_r \otimes\hod + \gamma \frac {\widetilde
  e^\perp}{r^2} \left(\id-P^\perp_r\right)\otimes \id
\end{equation}
for any $0\leq \gamma\leq 1$. If we choose $\gamma= (r^2/\widetilde
e^\perp)\eod$, the lowest $k$
eigenvalues of the operator on the right side of (\ref{finb}) are given by
$(1-\gamma)(1-\eta_{\rm L})E_{\rm 1d}^i(n,\ell,g)$,
$1\leq i \leq k$. This implies the lower
bound~(\ref{lowthm1}) of Theorem~\ref{thm1}.

\section{Estimates on Eigenfunctions}\label{sect:ef}

In this final section, we shall show how the estimates in the previous two
sections can be combined to yield a relation between the eigenfunctions
$\Psi_k$ of $\htd$ and the eigenfunctions $\psi_k$ of $\hod$. We start
with the following simple Lemma.

\begin{lem}\label{lem2}
  Let $H$ be a non-negative Hamiltonian on a Hilbertspace $\mathcal
  H$, with eigenvalues $0\leq E_1 \leq E_2 \leq \ldots$ and
  corresponding (orthonormal) eigenstates $\psi_i$.  For $k\geq 1$,
  let $f_i$, $1\leq i\leq k$, be orthonormal states in $\mathcal H$,
  with the property that $\langle f_i|H|f_i\rangle \leq \eta E_i$ for
  some $\eta > 1$.  If $E_{k+1}> E_k$, then
\[\sum_{i,j=1}^{k} |\langle f_i|\psi_j\rangle|^2 \geq
k-\frac{(\eta-1)\sum_{i=1}^{k}E_i}{E_{k+1}-E_k}\,.\]
\end{lem}

\begin{proof}
Let $P_k$ be the projection onto the first $k$ eigenfunctions of $H$. Then
$H \geq HP_k + E_{k+1} (1-P_k)$. By the variational principle, 
\[
\sum_{i=1}^k \langle f_i| HP_k + E_k (1-P_k) |f_i\rangle \geq
\sum_{i=1}^k E_i \,, \]
and hence
\[
\sum_{i=1}^k \langle f_i|H|f_i\rangle \geq \sum_{i=1}^k E_i +
(E_{k+1}-E_k)\sum_{i=1}^k \langle f_i| 1-P_k|f_i\rangle\,.\]
By assumption, $\langle f_i|H|f_i\rangle \leq \eta E_i$, implying the
desired bound 
\[
\sum_{i=1}^k \langle f_i| P_k |f_i\rangle \geq k
-\frac{(\eta-1)\sum_{i=1}^{k}E_i}{E_{k+1}-E_k}\,.\]
\end{proof}

\begin{cor}\label{maincor} Assume in addition that $E_{l+1}> E_l$ for some $1\leq l <
  k$. Then
\[
\sum_{i,j=l+1}^k |\langle f_i|\psi_j\rangle|^2 \geq k - l
-\frac{(\eta-1)\sum_{i=1}^{k}E_i}{E_{k+1}-E_k}
-\frac{(\eta-1)\sum_{i=1}^{l}E_i}{E_{l+1}-E_l}\,.\]
\end{cor}

\begin{proof}
Applying Lemma~\ref{lem2} and using the orthonormality of the $\psi_i$
and the $f_i$, we have
\begin{align}\nonumber
\sum_{i,j=l+1}^{k}|\langle f_i|\psi_j\rangle|^2 & =
\sum_{i,j=1}^{k}|\langle f_i|\psi_j\rangle|^2 +
\sum_{i,j=1}^{l}|\langle f_i|\psi_j\rangle|^2 \\ \nonumber & \quad - \sum_{i=1}^l
\sum_{j=1}^{k}|\langle f_i|\psi_j\rangle|^2 -  \sum_{i=1}^k
\sum_{j=1}^{l}|\langle f_i|\psi_j\rangle|^2 \\ & \geq k
-\frac{(\eta-1)\sum_{i=1}^{k}E_i}{E_{k+1}-E_k} + l
-\frac{(\eta-1)\sum_{i=1}^{l}E_i}{E_{l+1}-E_l}  - 2 l \,. \nonumber
\end{align}
\end{proof}

Let again $P^\perp_r$ denote the projection onto the function $\prod_{k=1}^n
b_r(\x^\perp_k) \in L^2(\R^{2n})$, and recall inequality
(\ref{finb}), which states that 
\begin{equation}\label{finba}
\htd  - \frac{ n e^\perp}{r^2} \geq (1-\gamma) \left(1-\eta_{\rm
    L}\right)  P^\perp_r \otimes \hod + \gamma \frac {\widetilde
  e^\perp}{r^2}  \left(\id-P^\perp_r\right) \otimes \id 
\end{equation}
for any $0\leq \gamma\leq 1$. As already noted, the choice  
$$\gamma= (r^2/\widetilde
e^\perp)\eod$$ implies that the lowest $k$
eigenvalues of the operator on the right side of (\ref{finba}) are given by
$(1-\gamma)(1-\eta_{\rm L})E_{\rm 1d}^i(n,\ell,g)$,
$1\leq i \leq k$. Here, we have to assume that 
 $\widetilde e^\perp / r^2 \geq \eod$ in order for $\gamma$ not to be
 greater than one. 
The corresponding eigenfunctions are 
\[
\psi_i(z_1,\dots,z_n) \prod_{k=1}^n b_r(\x^\perp_k)\,.
\]

We can now apply Corollary~\ref{maincor}, with $H$ equal to the
operator on the right side of (\ref{finba}), $f_i =
\Psi_i$, $l=k_i-1$
and $k=k_{i+1}-1$. Since $\etd \leq \eod /(1-\eta_{\rm U})$
by Theorem~1(b), we conclude that 
\begin{align}\nonumber
&\sum_{k=k_i}^{k_{i+1}-1}\sum_{l=k_i}^{k_{i+1}-1} \left| \left\langle \Psi_k \left| \psi_l
    \prod_{i=1}^n b_r(\x^\perp_k)\right. \right\rangle\right|^2 \\ &
\nonumber \geq
k_{i+1}-k_{i} - \left(\frac 1{(1-\gamma)(1-\eta_{\rm U})(1-\eta_{\rm
      L})}-1\right) 
\left[ \frac{\sum_{i=1}^{k_{i+1}-1}E_{\rm
    1d}^i}{E_{\rm 1d}^{k_{i+1}}-E_{\rm 1d}^{k_i}}+\frac{\sum_{i=1}^{k_i-1}E_{\rm
    1d}^i}{E_{\rm 1d}^{k_{i}}-E_{\rm 1d}^{k_i-1}}\right]
\end{align}
as long as $\gamma<1$, $\eta_{\rm U}<1$ and $\eta_{\rm L}<1$. 
Here, we used the notation $E_{\rm 1d}^i = E_{\rm 1d}^i(n,\ell,g)$ for
short. Since 
\[ \sum_{l=k_i}^{k_{i+1}-1} \left| \left\langle \Psi_k \left| \psi_l
    \prod_{i=1}^n b_r(\x^\perp_k)\right. \right\rangle\right|^2 
\leq 1
\]
for every  $k$, this implies Theorem~\ref{thm2}.

\bigskip

\textit{Acknowledgments}: We are grateful to Zhenqiu Xie for helpful
discussions. R.S. acknowledges partial support by U.S. NSF grant
PHY-0652356 and by an A.P. Sloan Fellowship.

\end{document}